

\input harvmac
\Title
{ \vbox
{ \hbox
{ RU-93-33}
\hbox { hep-th/9308039 } } }
{ { \vbox
{ \centerline
{ The Zamolodchikov $C$-Function, }
 \centerline
{ Classical Closed String Field Theory, }
\smallskip
\centerline
{ The Duistermaat-Heckman Theorem, }
\smallskip
\centerline
{ The Renormalization Group, and all that ...} } } }

\centerline { Kelly Jay Davis\footnote{${}^{\dagger}$}
{ Sponsored by Mom and Dad}}
\smallskip
\centerline { Department of  Physics and Astronomy }
\centerline { Rutgers University }
\centerline { Piscataway, NJ 08855 }
\vskip .5in

\centerline
{ \bf { Abstract } }

In this article we formulate a `topological' field theory by
employing a generalization of the Duistermaat-Heckman Theorem to
localize the path-integral of the `topological action' $C^2$ , where
$C$ is a slight modification of the Zamolodchikov $C$-Function, over
the space of all two-dimensional field theories to the fixed points
of the renormalization group's identity component. Also, we propose
an interpretation of the background independent classical closed
string field theory action $S$ in terms of the Zamolodchikov
$C$-Function's modification.

\Date { 7/25/93 }

\newsec { Introduction }

In this article we will formulate a `topological' field theory which has
the action $C^2$, $C$ being a slight modification of the Zamolodchikov
$C$-Function, by employing a generalization of the Duistermaat-Heckman
Theorem to localize the path-integral over the space of two-dimensional
field theories to the fixed points of the renormalization group.
Furthermore, we will express the background independent classical
closed string field theory action $S$ in terms of the Zamolodchikov
$C$-Function's modification. As far as the author knows, this is the first
example of an explicitly background independent classical closed string
field theory.

Previous formulations of closed string field theory have run up against
several seemingly impenetrable barriers. The first, and most significant,
barrier which these formulations have faced is that of background
independence. Basically, all such string field theories start by postulating
that space-time interactions are governed by a string theory which is
conformally invariant \ref \BZ { B. Zwiebach {\it Closed String Field
Theory: Quantum Action and the B-V Master Equation} Hep-Th/9206084}.
However, requiring such a theory to be conformally invariant
is equivalent to requiring that the background space-time fields
are solutions to string theory's classical equations of motion. Thus, such
string field theories are said to be background dependent, as they depend
upon a classical background. String field theory then makes its appearance
only to describe the perturbations of these fields about their classical
backgrounds. This state of affairs is unsatisfactory on at least two fronts.
First of all, the string theorist, not the string theory, chooses the
particular classical background. Thus, any relevant physics which
string theory may have graced us with is lost. Secondly, the interactions
are only formulated perturbatively. Any non-perturbative interactions,
which, for instance, would act to choose the vacuum, are ignored.
Therefore, this state of affairs seems to desperately call for a background
independent notion of string field theory. In this paper we will attempt to
take the first steps towards this goal by formulating a background
independent classical closed string field theory.

The second problem which string field theory faces is that of obtaining a
well defined configuration space. Conventionally, one defines the
configuration space of a theory to be the set of all possible space-time fields
modded by an equivalence relation derived from the theory's gauge group.
However, in contrast to most `conventional' field theories, the space-time
fields in closed string theory appear in a very unusual setting.
In the case of closed string field theory, the set of space-time fields do not
appear as dynamical fields in the two-dimensional field theory, they
appear as coupling constants. Thus, if we consider a two-dimensional
field theory as being completely defined by a set of coupling constants, the
configuration space of string field theory can be taken to be the space of
`all' two-dimensional field theories modded by some appropriate gauge
group. However, there are various problems which arise with such a
simplistic outlook. Ideally, to properly formulate a closed string field
theory one would like to impose various structures on the space of `all'
two-dimensional field theories such as a metric, a symplectic form, a
volume form \dots\ But, to impose such structures on the space of `all'
two-dimensional field theories is a difficult task. All such attempts have
settled with rather badly defined structures over this space \ref \WA {
E. Witten {\it On Background Independent Open-String Field Theory}
Hep-Th/9208072}. In this article we will avoid the most difficult
of these problems by simply assuming there exists a symplectic
form over the space of `all' two-dimensional field theories.

A third hurdle which string field theory has faced is the search for a
formulation of a full, background independent classical closed string
field theory action. The major obstacle to this goal is our lack of knowledge
in two areas: formulating string field theories around backgrounds which
are not solutions to string theory's classical equations of motion and
formulating string field theories around backgrounds which correspond
to non-renormalizable two-dimensional field theories {\BZ \WA}.
However, again, in this paper we side-step this problem by only considering
very general properties of the background independent classical closed
string field theory action which do not depend upon the backgrounds being
solutions to closed string field theory's classical equations of
motion or corresponding to renormalizable two-dimensional fields theories.

The outlay of this paper is as follows. First, we will review a non-abelian
generalization of the Duistermaat-Heckman Theorem, this will occupy
the second section. Second, we will define the configuration space of our
theory and some of its geometrical properties, this will occupy section
three. After this, we will set out to define the renormalization group of our
theory and some of its topological properties. Finally, in the fifth
section, we will tie this all together and compute some path-integrals in
the theory $C^2$ and derive the relation between the background independent
classical closed string field theory action $S$ and the Zamolodchikov
$C$-Function's modification. The sixth section will be occupied with
conclusions
and wild speculations.

\newsec { The Non-Abelian Duistermaat-Heckman Theorem }

In this section we will explain the non-abelian version of the
Duistermaat-Heckman Theorem, following the work of Witten
\ref \WB { E. Witten {\it Two Dimensional Gauge Theories
Revisited} Hep-Th/9204083 }. We will explain the theorem
in two steps. The first step will occupy the first subsection, and
the second step will occupy the second subsection. In the first
subsection we will explain the notion of equivariant integration,
and in the second subsection we will explain the localization
principle itself.

\subsec { Equivariant Integration }

In this subsection we will explain equivariant integration. This
explanation will consist of two portions. In the first portion we will
introduce equivariant cohomology, and in the second portion we will
introduce equivariant integration.

Let us start by considering a manifold $X$ which is acted upon by
a compact, connected group $G$ with Lie algebra ${\cal G}$.
Furthermore, let us assume that the manifold $X$ is symplectic
and of dimension $2n$. Now, consider the deRham complex of $X$ with
complex coefficients, $\Omega^{*}(X)$, and the space of zero-forms on
${\cal G}$, $\Lambda^{0}({\cal G})$. Also, let us grade $\Lambda^{0}
({\cal G})$ such that a $n^{th}$ order homogenous polynomial is of
degree $2n$. Now, we define the equivariant forms on $X$ to be the
elements of $\Omega^{*}(X) \otimes \Lambda^{0}({\cal G})$ which are
invariant under the action of $G$. Let us denote the equivariant
differential forms over $X$ by $\Omega^{*}{}_{G}(X)$.

As we have a notion of an equivariant differential form, let us now
consider defining a notion of equivariant cohomology. With this goal
in mind, we must first define an equivariant deRham $d$ operator. This
is done by considering the $G$ action on $X$. The $G$ action on $X$ is
given by a homomorphisim from an element in ${\cal G}$ to a vector field
on $X$, and the flow along this field is $G$'s action. Thus, if $v$ is an
element of ${\cal G}$, then there is a corresponding vector field $V(v)$ on
$X$.  Then, we define the equivariant deRham $d$ operator $d_{G}$ as,

$$
d_{G} \equiv
d - i i_{V(v)},
\eqno (2.1.1)
$$
\vskip 5mm

\noindent where $i_{V(v)}$ denotes contraction with the vector field
$V(v)$. However, a short computation yields the following result,

$$
d^{2}_{G} =
-i {\cal L}_{V(v)},
\eqno (2.1.2)
$$
\vskip 5mm

\noindent where ${\cal L}_{V(v)}$ denotes Lie differentiation with
respect to the vector field $V(v)$. Thus, the operator is nil-potent
on precisely the equivariant differential forms. Therefore, we have
a natural notion of a $G$-equivariant cohomology precisely on
the equivariant differential forms. Let us denote this $G$-equivariant
cohomology by $H^{*}_{G}(X)$.

As we have now defined the notion of a $G$-equivariant cohomology, let
us define the notion of equivariant integration. Among the things
on our wish list for the properties of equivariant integration, we should
request that the integration only depend upon the equivariant
cohomology class of the integrand, and, of course, the integral should
not diverge. These two points will guide our definition.

The vector space ${\cal G}$ has a natural invariant measure, unique
up to a constant factor. To fix this factor let us consider ${\cal G}$ in
a different setting; ${\cal G}$ is, by definition, in isomorphism with
$TG|_{id}$.  A choice of Harr measure on $G$ then defines a measure
on ${\cal G}$. As $G$ is compact the Harr measure on $G$ yields a
finite volume $Vol(G)$ for $G$. Now, choose coordinates $v_m$ on
${\cal G}$ such that the measure $dv_1 dv_2 \dots dv_{m}$ on ${\cal G}$
coincides with the Harr measure at $id$ of $G$. Thus, we now have a
natural measure,

$$
{ { 1 } \over { Vol(G) } }
dv_1 dv_2 \dots dv_m,
\eqno (2.1.3)
$$
\vskip 5mm

\noindent on ${\cal G}$ which is independent of the chosen Harr
measure. The equivariant integration we wish to define on $X$ is now
taken to be the map from $H^{*}_{G}(X) \rightarrow \Re$  given by,

$$
\alpha
\rightarrow
{ { 1 } \over { Vol(G) } }
\int_{ {\cal G} \times {X} }
{ { dv_1 dv_2 \dots dv_m } \over { {(2 \pi)}^m } }~
\alpha.
\eqno (2.1.4)
$$
\vskip 5mm

\noindent However, this definition is not quite up to snuff as the integral
does not generically converge. But, we may fix this by putting in a
convergence factor. If we take $s$ as a positive, real number and $(~,~)$
as a positive definite invariant quadratic form on ${\cal G}$, then we
define equivariant integration of an element $\alpha$ in $H^{*}_{G}(X)$
as follows,

$$
\oint_{X} \alpha \equiv
{ { 1 } \over {Vol(G)} }
\int_{ {\cal G} \times {X} }
{ { dv_1 dv_2 \dots dv_m } \over { {(2 \pi)}^m } }~
\alpha
\exp
\Big(
- { { 1 } \over {4 s} }
(v,v)
\Big).
\eqno (2.1.5)
$$
\vskip 5mm

\noindent With this added exponential convergence factor, we can
integrate equivariant differential forms with arbitrary polynomial
dependence upon $v$ and all such integrals will converge as a result
of our exponential factor. We will call the above map equivariant
integration.

\subsec { Localization Principle }

In this subsection we will explain the non-abelian localization principle
which is a generalization of the localization principle of Duistermaat
and Heckman.

Consider an equivariantly closed form $\alpha$ on $X$; then, for any real
number $t$ and any `nice' $\lambda~\epsilon~\Omega^{*}_{G}(X)$, one
has,

$$
\oint_{X} \alpha =
\oint_{X} \alpha
\exp
\Big(
t d_{G} \lambda
\Big).
\eqno (2.2.1)
$$
\vskip 5mm

\noindent This is a result of the fact that the form $\alpha ( 1 - \exp (
t d_{G} \lambda ) )$ is equivariantly exact and thus integrates to zero,
by construction. Thus, if one writes the integral of $\alpha ( 1 - \exp (
t d_{G} \lambda ) )$ and takes the $\alpha$ term to one side and the
$\exp$ term to the other side, we have the result.

Now, let us consider a specific case of the above formula. For our purposes,
we will consider $\alpha$ to be independent of $v$ and we shall assume
$\lambda$ is independent of $v$ also. Furthermore, we shall assume that
$\lambda$ is an equivariant one-form. Thus, if we choose an orthonormal
basis $T_a$ of ${\cal G}$ and write $V(v)$ as $V_av^a$, where $V_a$ is a
vector field on $X$ corresponding to $T_a$ and $v^a$ are linear
functions on ${\cal G}$. Then, equation (2.2.1) takes the form,

$$
\oint_{X} \alpha =
{ { 1 } \over { Vol(G) } }
\int_{ {\cal G} \times X}
{ { dv_1 dv_2 \dots dv_m } \over { {(2 \pi)}^m } }~
\alpha
\exp
\Bigg(
t d \lambda -
it \lambda(V_a) v^a -
{ { 1 } \over {4 s} }  (v^a,v^a)
\Bigg),
\eqno (2.2.2)
$$
\vskip 5mm

\noindent where repeated indices are summed over. As the only $v^a$
dependence resides in the exponential, we may complete the square
and integrate out the $v_a$ dependence. Thus, one obtains,

$$
\oint_{X}
\alpha =
{ { 1 } \over { Vol(G) {(\pi / s)}^{m / 2} } }
\int_{X}
\alpha
\exp
\Bigg(
t d \lambda -
t^{2} s ( \lambda(V_a), \lambda(V_a) )
\Bigg).
\eqno (2.2.3)
$$
\vskip 5mm

Now, as the integral is formally independent of the value of $t$, we may
take the limit as $t \rightarrow \infty$. Thus, upon taking this limit, one
sees that the points on $X$ at which $\lambda (V_a) \not= 0$ do not
contribute to the above integral. Thus, the integral is localized to the
set of points on $X$ at which $\lambda (V_a) = 0$. Let us enumerate
the connected components of this set  by the index $\sigma~\epsilon~U$,
where $U$ is some indexing set. Thus, the integral above takes the form,

$$
\oint_{X} \alpha =
\sum_{ \sigma~\epsilon~U }
Z_{ \sigma },
\eqno (2.2.4)
$$
\vskip 5mm

\noindent where the summand $Z_{ \sigma }$ corresponds to the
contribution given by the connected component $\sigma$. Now, let us
consider a particular example.

If the action of $G$ on $X$ is Hamiltonian, as we assume it is, then
corresponding to $V_a$ there exists a function on $X$, $\mu_a$ say, such
that $-i_{V_a} \omega= d \mu_a$, where $\omega$ is the symplectic
form on $X$. Now, let us consider an $\alpha$ for our specific example
given by,

$$
\alpha =
\exp ( \omega - i\mu_a v^a ).
\eqno (2.2.5)
$$
\vskip 5mm

\noindent Thus, with this particular $\alpha$ the equivariant integral
takes the form,

$$
\oint_{X} \alpha =
{ { 1 } \over { Vol(G) } }
\int_{{\cal G} \times X}
{ { dv_1 dv_2 \dots dv_m } \over { {(2 \pi)}^{m} } }
\exp
\Bigg(
\omega -
i\mu_a v^a -
{ { 1 } \over { 4s } } (v^a,v^a)
\Bigg).
\eqno (2.2.6)
$$
\vskip 5mm

\noindent Now, if we perform the $v^a$ integral, then the above integral
takes the form,

$$
\oint_{X} \alpha =
{ { 1 } \over { Vol(G) {(\pi / s)}^{ {m} \over {2} } } }
\int_X
{ {\omega^n} \over {n!} }
\exp
\Big(
-s (\mu , \mu)
\Big).
\eqno (2.2.7)
$$
\vskip 5mm

\noindent Thus, in our particular case, the integral is localized over the
elements of $X$ at which $( \mu, \mu ) = 0$. This result will come in
handy later on when we apply all this stuff to background independent
classical closed string field theory and the Zamolodchikov $C$-Function.
In fact, this will be the exact integral that we employ.

\newsec { The Space of All Two-Dimensional Field Theories }

In this section we will derive the basic properties of the space of all
two-dimensional field theories. First, let us consider what we will mean
when we say the space of `all' two-dimensional field theories. Essentially,
what we will mean is the space of all two-dimensional field theories which
have an interpretation as string theories. Therefore, if we wish to consider
all such field theories, we must have a natural notion of what fields are
space-time fields and what fields are world-sheet fields. This notion
is easily obtained by limiting ourselves in considering only the common
world-sheet fields $X^{ \rho }$ and $h_{ab}$. Given these world-sheet
fields, we may now easily construct the space-time fields of our theory.
This is done by first considering all possible combinations of the
world-sheet fields $\{ X^{ \rho },h_{ ab }, { \epsilon }_{ab}, { \nabla }_{ a_{
1} } X^{ \rho }, { \nabla }_{ a_{ 2 } } { \nabla }_{ a_{ 1} } X^{ \rho }, \dots
\}$
into operators with only space-time indices. For example, one could have
an operator of the form $h^{ab} { \nabla }_{ a } X^{ \rho } { \nabla }_{ b }
X^{ \xi } \sqrt { h }$. Now, let us denote an arbitrary such operator as ${
\bar O } ^{ { \rho }_1 { \rho }_2 \dots { \rho_n } }(r)$, where $r$ is a
world-sheet point. Given the set of all such operators, let us introduce one
space-time field for each such operator. For instance, ${ \bar O }^{ { \rho
}_1 { \rho }_2 \dots { \rho_n } }(r)$ would correspond to a space-time field
${\bar F }_{ { \rho }_1 { \rho }_2 \dots {\rho_n }}(x)$, where $x$ is a
space-time point. Now, with this information we may define a
two-dimensional field theory.

We can define a two-dimensional field theory with this data by simply
defining the Lagrangian to be the summation of all possible operators
contracted with their corresponding space-time fields. More specifically,
we define the action of this two-dimensional theory as follows,

$$
S_{2DFT} = \int_{ \Sigma }
\sum_{ { All Possible } \atop
{ { \bar O } ^{ { \rho }_1 \dots { \rho }_n } (r) } }
{ \bar O } ^{ { \rho }_1 { \rho }_2 \dots { \rho_n } } ( r )
{ \bar F }_{ { \rho }_1 { \rho }_2 \dots { \rho_n } } ( X^{ \rho } ( r ) )~
d^2r,
\eqno (3.1)
$$
\vskip 5mm

\noindent
where $\Sigma$ is a Riemann surface. However, note that this is a
two-dimensional field theory and the dynamical fields are $X^{ \rho }$
and $h_{ab}$, not the space-time fields. The space-time fields only
appear as coupling constants. Thus, any given set of space-time fields
which saturates all possible ${ \bar O }^{ { \rho }_1 { \rho }_2 \dots
{ \rho_n } }( r )$'s defines a two-dimensional field theory for $X^{ \rho }$
and $h_{ab}$. Therefore, the space of all two-dimensional field theories
is equivalent to the space of all possible sets of space-time fields which
saturate the ${ \bar O }^{ { \rho }_1 { \rho }_2 \dots { \rho_n } } ( r )$'s.

However, in our considerations we will only be concerned with the space
of all renormalizable two-dimensional interactions. Therefore, we will
only have need of the renormalizable interactions in equation (3.1). Let us
denote a generic renormalizable operator by $O^{ { \rho }_1 { \rho }_2 \dots
{ \rho_n } } ( r )$. Similarly, let us denote a space-time field which
corresponds to a two-dimensional renormalizable operator by $F_{
{ \rho }_1 { \rho }_2 \dots { \rho_n } } ( x )$.  Thus, as before, the space of
all renormalizable two-dimensional field theories is equivalent to the space
of all possible sets of space-time fields which saturate the $O^{ { \rho }_1
{ \rho }_2 \dots { \rho_n } } ( r )$'s. Let us denote the space of all
renormalizable two-dimensional field theories by ${ \bar { \cal M } }$. A
point in this space is a set of space-time fields $ \{ \dots, F_{ { \rho }_1
{ \rho }_2 \dots { \rho_n } }(x), F_{ { \rho }_1 { \rho }_2 \dots { \rho_m } }
(x), \dots \} $ which saturates all possible renormalizable combinations of
$\{ X^{ \rho },h_{ ab }, { \epsilon }_{ab}, { \nabla }_{ a_{ 1} } X^{ \rho },
{ \nabla }_{ a_{ 2 } } { \nabla }_{ a_{ 1} } X^{ \rho }, \dots \} $.

Now, as in most cases of `theories with symmetries,' we must mod the
na\"{ \i }ve configuration space of the theory, ${ \bar { \cal M } }$ in this
case, by the gauge symmetries of the theory. However, we run into a bit of
trouble here as the fields in our na\"{\i}ve configuration space ${ \bar
{ \cal M } }$ do not appear dynamically in the two-dimensional action,
they appear as coupling constants. But, we have no `natural' notion
of a gauge symmetry for coupling constants. Thus, we must improvise. In
general, a gauge symmetry is a canonical transformation of the fields in the
theory. A canonical transformation on the theory's fields is defined in such
a manner that it leaves the Lagrangian invariant up to a total
derivative.Thus, if we extend this notion to the case at hand, we can define a
gauge transformation of $p~\epsilon~{ \bar { \cal M } }$ to be any
transformation of $p$'s fields which leaves the two-dimensional theory $p$
corresponds to invariant up to a total derivative. As one may check
explicitly in the case of the space-time metric, which corresponds to the
operator $h^{ab}{ \nabla }_{ a } X^{ \rho } { \nabla }_{ b } X^{ \xi } \sqrt
{ h }$, this is the correct prescription. Thus, as we now have a natural
notion of a gauge symmetry for all the fields in ${ \bar { \cal M } }$, we
are ready to pass to the true configuration space. This is simply done by
modding ${ \bar { \cal M } }$ by the action of the gauge symmetries above.
Let us denote the resulting space by ${ \cal M }$, the true configuration
space.

As we now have some knowledge as to what the true configuration
space of closed string field theory is as a set, we may further explore
its properties by imposing various structures on ${\cal M}$, a metric
and a symplectic form.

First, we will impose a metric upon the space ${\cal M}$. Before we set
off demanding a metric on ${\cal M}$, let us remind ourselves of what a
metric is. A metric at a point $p~\epsilon~{\cal M}$, by definition, is a
multi-linear symmetric non-degenerate map from $T{\cal M} \otimes
T{\cal M}|_p$ to the real numbers. Thus, if we are to define a metric on
${\cal M}$, we must exhibit such a map for all $p~\epsilon~{\cal M}$.
Let us choose some $p~\epsilon~{\cal M}$ and also two elements of $T{
\cal M}|_p$, $V_{1}{}_{\cal A}$ and $V_{2}{}_{\cal B}$ say, where the
indices ${\cal A}$ and ${\cal B}$ run over all of the `field directions' of
${\cal M}$. Now, we must exhibit a scalar corresponding to these two
vectors. Let us first note that we have a natural map from the space of
vectors over $p$ to the space of operator valued vectors over $p$. This
is simply the map $V_{\cal A} \rightarrow O^{\cal A}V_{\cal A}$.
Now, we have a natural notion of a scalar which is associated to two
operator valued vectors, their space-time scattering amplitude. Thus, we
define the metric $g^{\cal AB}$ on ${\cal M}$ at $p$ as follows,

$$
g^{\cal AB} V_{1}{}_{\cal A} V_{2}{}_{\cal B} \equiv
\sum_{ { {All Riemann} \atop {Surfaces \Sigma Mod} }
\atop {Symmetries} }
\int_{\Sigma \times \Sigma}
< O^{\cal A} V_{1}{}_{\cal A}(\sigma_1)
O^{\cal B} V_{2}{}_{\cal B}(\sigma_2) >~
d^2 \sigma_1~
d^2 \sigma_2,
\eqno (3.2)
$$
\vskip 5mm

\noindent where the summation is taken over all $g=0$ Riemann surfaces mod
the symmetries of the two-dimensional theory defined by the point $p$.
As one may note, this is a rather poorly defined metric, some of its entries
may yield infinities. However, to get rid of these infinities, we will
introduce a world-sheet cut-off $\lambda$. With this cut-off in
place, the metric above is well defined, but dependent upon the cut-off.
However, in the next section we will implement a renormalization program
which will allow us to remove the dependence upon the cut-off. Also,
note that the metric is trivially multi-linear and symmetric. Its
non-degeneracy follows from its relation to Zamolodchikov's metric
\ref \AZ { A. Zamolodchikov {\it `Irreversibility' of the Flux of the
Renormalization Group in a 2D Field Theory } {\bf JETP Letters } 43 ( 1986 )
730-732 } or from space-time unitary.

Now, we will simply assume the existence of a symplectic form $\omega_
{\cal AB}$ over ${\cal M}$. As seen in Witten's paper {\WA}, the construction
of such a symplectic form is fiendishly difficult in the case of open strings,
and, as a rule, closed strings present much more of a problem. Therefore,
we will simply assume one exists and see what we may derive from its
existence. With these two structures over ${\cal M}$, we are ready to use
some of ${\cal M}$'s geometrical properties to define the renormalization
group and explore the renormalization group's topological properties.

\newsec { The Renormalization Group }

In this section we will derive various properties of the renormalization
group which acts on ${ \cal M }$, the true configuration space. This section
will be divided into two subsections. In the first subsection we will loosely
recount the article of Dole \ref \BD { B. Dolan{ \it The Renormalization
Group Equation as an Equation for Lie Transport of Amplitudes }
Hep-Th/9307024 } on the renormalization `group.' In the second subsection
we will define a more general notion of the renormalization group, the
actual group as opposed to the renormalization Lie algebra which
is commonly refered to as the renormalization group. Then, we will
derive various topological properties of the renormalization group
which we will use in our computations.

\subsec { Renormalization Lie Algebra and Lie Transport }

In this subsection we will derive various properties of the
renormalization Lie algebra, which is commonly called the
renormalization group. These geometrical properties of the
renormalization Lie algebra will serve as motivation for the
definition of the full renormalization group in the next subsection.

Let us start this subsection by noting that, as mentioned previously,
the space-time fields, which define coordinates on the space ${ \cal
M }$, appear as coupling constants in the two-dimensional
world-sheet field theories. However, as of yet, we have not
renormalized these parameters. Thus, they are bare parameters and,
as of yet, unphysical. Thus, as in conventional quantum field theory,
we must implement some renormalization program so as to yield
physical values for the coordinates of ${\cal M }$.

Let us start in a rather conventional manner by assuming that
the bare fields which describe points in ${ \cal M }$ are
functions of the renormalized fields. Also, let us introduce a
regularization parameter or parameters $\epsilon$; for a cut-off
$\lambda$ one has $\epsilon = \kappa / \lambda$, where $\kappa$
is a renormalization point. Furthermore, let us assume the bare
fields are functions of the regularization parameter(s) $\epsilon$.
Now, as the bare couplings previously provided a
coordinate system on the space ${ \cal M }$, the new
renormalized couplings will also provide a coordinate system on
${ \cal M }$, provided that the transformation between the bare
and renormalized couplings is not singular. Similarly, if we
choose some other renormalization scheme, this will lead to a
new set of renormalized space-time fields and thus a new set of
coordinates on ${ \cal M }$. In other words, various schemes of
renormalization correspond to different coordinate systems on
${ \cal M }$.

Now, let us consider how correlation functions are affected by
the choice of renormalization scheme. Let us consider computing
a correlation function, in some renormalization scheme, which
is of the following form $< \dots O_{ N } (r_n) O_{ M }(r_m) \dots >$,
where $r_n$ and $r_m$ are points on a Riemann surface such that
$r_n \not= r_m$ for all $n \not= m$ and the indices
$N$ and $M$ refer to specific operators in the space of
renormalizable two-dimensional operators. If we choose some
second renormalization scheme in which to compute this
correlation function, we find a new correlation function. As a new
renormalization scheme is equivalent to a change of coordinates
on ${ \cal M }$, the new coordinates of ${ \cal M }$ will be related
to the old via a simple functional relationship. In other words,
if the old renormalization scheme had coordinates $F^{ \cal A }$,
where ${\cal A}$ runs over all `field directions' of the theory, the new
scheme has coordinates ${\hat F}^{\cal A}(F^{\cal A})$.
The correlation function, as computed in the new system, is then
related to the correlattion function computed in the
old system via a tensor transformation {\BD} \ref \ML { M. L\"assig
{\it Geometry of the Renormalization Group with an Application
in Two-Dimensions} {\bf Nuclear Physics B} 334 ( 1990 ) 652-668 }.
In other words, such correlation functions are tensors over
${\cal M}$ and transform accordingly under a change of
coordinates on ${\cal M}$, which, of course, corresponds to a
change in the renormalization scheme.

However, as we wish physically measurable correlation functions
to be invariant under choice of renormalization scheme, such
correlation functions are not to be considered `physical.' We
need to define some other quantity which holds all of the
information of these correlation functions, but is invariant
under a change of coordinates on ${\cal M}$, equivalently,
a change in the renormalization scheme. Thus, what we really
desire is to describe the correlation functions as scalars on ${
\cal M}$. The simplest manner in which to get a scalar from the
tensor correlation functions above is to contract the tensors above
with the appropriate number of one-forms so as to obtain a
scalar. This is what we shall do.

If we consider a point $p$ which is in ${ \cal M }$, then it is
defined by a set of space-time fields $ \{ \dots, F^{ N }(x), F^{ M } (x),
\dots \} $ which saturates all possible renormalizable combinations
of $\{ X^{ \rho },h_{ ab }, { \epsilon }_{ab}, { \nabla }_{ a_{ 1} }
X^{ \rho }, { \nabla }_{ a_{ 2 } } { \nabla }_{ a_{ 1} } X^{ \rho },
\dots \} $. Thus, a one-form on ${\cal M}$ at $p$ corresponds to an
arbitrary variation of the fields which define $p$ modded by the
gauge symmetries of the theory. In other words,
$dF^{ N } = \delta F^{ N }$ mod symmetries. Thus, if we have a
correlation function of the form $< \dots O_{ N }(r_n) O_{ M }(r_m)
\dots >$, then we must produce a set of one-forms $\{ \dots dF^{ N },
dF^{ M } \dots \}$. With these one-forms we can create a scalar
over ${ \cal M }$ by contracting, $< \dots O_{ N } dF^{ N }(r_n)
O_{ M } dF^{ M }(r_m) \dots >$. This quantity is
a scalar over ${ \cal  M}$ and is therefore invariant under
changes in the coordinates on ${ \cal M }$, which, of course,
correspond to changes in the renormalization scheme. Therefore,
we will take these scalars as our physical observables.

However, the thrust of renormalization theory is the idea that the
physical observables calculated at some point $p~\epsilon~{\cal M}$
should be the same as those calculated at a point $p+\delta \beta$,
where $\delta$ is an infinitesimal parameter and $\beta$ is the theory's
beta-function. In other words, the beta-function $\beta_{\cal A}$
is a vector field on ${\cal M}$ and it produces a one-parameter
flow on ${\cal M}$ such that the physical correlation functions are
invariant under this flow. In terms of equations, we may write this as
follows,

$$
< \dots O_{ N } dF^{ N }(r_n)
O_{ M } dF^{ M }(r_m) \dots >(p) -
\eqno (4.1.1)
$$
$$
< \dots O_{ N } dF^{ N }(r_n)
O_{ M } dF^{ M }(r_m) \dots >(p+\delta \beta) =0.
$$
\vskip 5mm

Now, with this statement of the renormalization group equation, we can,
with a bit of mathematics, interpret the renormalization Lie algebra in a
more geometric setting. The first step in doing so is dividing both sides
of the above equation by $\delta$. Then, we take the limit as $\delta
\rightarrow 0$. Then, after noting the definition of the Lie derivative,
one obtains,

$$
{\cal L}_{\cal B}
< \dots O_{ N } dF^{ N }(r_n)
O_{ M } dF^{ M }(r_m) \dots >=0
\eqno (4.1.2)
$$
\vskip 5mm

\noindent In other words, the renormalization group equation
expresses the fact that the physical correlation functions of the
theory are invariant with respect to Lie transport along the vector
fields defined by the beta-functions.

\subsec { The Renormalization Group }

Before we plunge straight into a definition of the renormalization
group, let us consider a simple example from general relativity
which will act to guide us through this definition. In general
relativity when we consider a space-time manifold $M$ and the
space of diffeomorphisims which act on this manifold $Diff(M)$,
the tangent space at the identity of $Diff(M)$ is the Lie algebra
$diff(M)$. The Lie algebra $diff(M)$ acts on $M$ in a simple
manner. If $v$ is an element of this Lie algebra, then its action
on $M$ corresponds to a diffeomorphisim of $M$ generated by
a vector field $V(v)$ on $M$. This is simply the normal notion
of an infinitesimal diffeomorphisim, common in general
relativity. Let us now consider how this applies to our present
case of the renormalization group.

As we saw in the last section, the beta-functions act on
${ \cal M }$ as infinitesimal transformations. This is exactly
analogous to the action of the vector fields $V(v)$ on $M$ in
the case of general relativity. Thus, we are led to consider the
beta-functions as the vectors on ${ \cal M }$ which correspond
to the Lie algebra of the renormalization group. However, we
are now left with the task of defining exactly what the
renormalization group is.

If we consider the physical motivation behind the
renormalization program, then we see that its core idea is
that the physical correlation functions calculated at some point
$p~\epsilon~{\cal M}$ should be the same as those
calculated at a point $p+\delta \beta$. However, in the
last section we proved that this is equivalent
to requiring that the transformations of the renormalization Lie
algebra on ${ \cal M }$ be such that they are Killing vectors of all
physical correlation functions over ${ \cal M}$. Thus, with this
as a motivation, we should define the renormalization group
as the set of all homeomorphisims of ${ \cal M }$ to itself which
leave the physical correlation functions
invariant. In other words, if $\phi$ is a member of the
renormalization group ${ \cal RG }$, then for any physical
correlation function $< \dots O_{ N } dF^{ N }(r_n) O_{ M } dF^{ M }
(r_m)\dots >(p)$ at $p~\epsilon~{ \cal M }$ one has $< \dots O_{ N }
dF^{ N }(r_n) O_{ M } dF^{ M }(r_m) \dots >(p) = < \dots O_{ N }
dF^{ N }(r_n) O_{ M } dF^{ M }(r_m) \dots >( \phi (p))$. We
will take this as our definition of the renormalization group.
It seems the only logical definition which we may make, and,
of course, it has the correct tangent space, the beta-functions.

Now, as we've defined the renormalization group, let us
consider some of its topological properties. First of all, we
may introduce a `natural' metric $G(~,~)$ on ${ \cal RG }$.
To do this let us first consider what a metric on ${ \cal RG }$
would be. By definition, a metric at $q~\epsilon~{ \cal RG }$
is a multi-linear symmetric non-degenerate map from
$T{ \cal RG } \otimes T{ \cal RG }|_q$ to the real numbers.
Now, as we have previously said in the case that $q$ is the identity,
the tangent space of ${ \cal RG }$ at a point $q$ can be identified
with a set of vectors on $q({\cal M})$. Thus, a metric at $q$ on
${ \cal RG }$ is equivalent to a metric on the space of
vectors over $q({\cal M})$. To wit, if we consider two
elements of $T{ \cal RG }|_q$, $b_1$ and $b_2$ say,
then they correspond to two vector fields $\beta_{1}{}_
{ \cal A}$ and $\beta_2{}_{ \cal B }$ over $q({ \cal M })$.
Now, we may introduce a scalar which corresponds
to the inner product of $b_1$ and $b_2$,

$$
G(b_1,b_2) \equiv
\int_{q({ \cal M })} g^{ \cal AB }
\beta_{1}{}_{ \cal A }
\beta_{2}{}_{ \cal B }~
DF.
\eqno (4.2.1)
$$
\vskip 5mm

However,  as ${ \cal RG }$ is now a metric space, we can
appeal to a theorem of topology to draw some conclusions
about the behavior of ${ \cal RG }$. A standard theorem
of topology states that if a space is a metric space, then limit
point compactness is equivalent the to normal notion of
compactness. What this means is that to prove a metric
space is compact one only needs to prove that every infinite
sequence of points in the space has a limit point which is
also in the space. We will use this to explicitly compactify
${ \cal RG }$.

Consider the set of all infinite sequences in ${ \cal RG }$.
Let us denote this set by $ \{  \{ q_n \} \} $. Furthermore,
let us consider the set of all limit points of these sequences.
We will notate this set by $ \{ q_\infty \} $. Now, we will,
rather explicitly, compactify ${ \cal RG }$ by defining all
the points of $ \{ q_\infty \} $ to be elements of ${ \cal RG }$.
We will refer to this compactified renormalization group
as ${ \overline { \cal RG } } $. For future reference, we will
also need to consider the component of ${ \overline { \cal
RG } } $ which is homotopic to the identity. We will
call this space ${ \overline { \cal RG } }_0 $. Now that we've
explored some of the topological properties of the
renormalization group, we are ready to put them to use
in computing some exact path-integrals.

\newsec { String Field Theory and the Zamolodchikov $C$-Function}

In this section we will employ a non-abelian generalization
of the Duistermaat-Heckman Theorem to localize the path-integral
of the action $C^2$ over ${ \cal M }$ to the fixed points of the
group ${ \overline { \cal RG } }_0 $. Thus, as we will use the
non-abelian generalization of the Duistermaat-Heckman
Theorem, let us first remind ourselves of the arena in which
it applies.

Consider the action of a compact, connected group
$G$, with Lie algebra ${ \cal G }$, on a manifold $X$.
Let us assume that $X$ is a symplectic manifold
of dimension $2n$ with a symplectic form $\omega$.
The action of $G$ is said to be Hamiltonian, as we assume
it is, if it is induced from a map ${ \tilde \mu } : { \cal G }
\rightarrow\Lambda^0 (X)$. In other words, for each element
$t$ of $\cal G$ there exists a vector field $T(t)$ on $X$, which
represents the action of $t$ on $X$, and function $\mu
(t)$ on $X$ such that $-i_{T(t)} \omega = d \mu (t)$. As
$G$ is taken to be multi-dimensional, we can choose
various such $t$'s. Let us denote a set of them by $t_n$.
As the $t_n$ are members of ${ \cal G }$, a vector space,
we may take their vector sum $t_\Sigma$ and introduce a
function $\mu ( t_\Sigma )$ which corresponds to this
vector sum. Also, as the elements of $\cal G$
naturally correspond to vector fields on $X$, the dual
of ${\cal G}$, ${{ \cal G }^*}$, corresponds naturally
to some set of one-forms on $X$. Thus, we may associate to
the function $\mu(t_\Sigma )$ a natural element of
${{ \cal G }^*}$ given by $d\mu(t_\Sigma )$. Now, if we
introduce a positive definite invariant quadratic form $(~,~)$
on ${{ \cal G }^*}$, then we may consider integrals of the
form,

$$
Z =
\int_{X}
{ {\omega^n} \over {n!} }
\exp
\Big(
-s \big( d\mu(t_\Sigma ), d\mu(t_\Sigma) \big)
\Big),
\eqno (5.1)
$$
\vskip 5mm

\noindent where $s$ is a positive constant. Such
integrals can be exactly evaluated. As we proved previously,
they are localized about the minima of $\big( d\mu(t_\Sigma),
d\mu(t_\Sigma) \big)$. As $(~,~)$ is a positive definite invariant
quadratic form, the minima of $\big(d\mu(t_\Sigma), d\mu(t_
\Sigma) \big)$ occur at the points $d\mu(t_\Sigma) = 0$. Thus,
the integral is equivalent to a summation of various terms, each
term in the sum corresponds to a connected component
of the set of points at which $d\mu(t_\Sigma) = 0$.
Thus, the integral takes the following form,

$$
Z =
\sum_{ {Components} \atop { of d\mu(t_\Sigma) = 0} }
Z_{ \sigma }.
\eqno (5.2)
$$
\vskip 5mm

Now, let us consider how to apply this to the case at hand. In
the case at hand, we will take the manifold $X$ as ${\cal M}$.
However, we will not take the group $G$ as ${\overline {\cal RG}
}_{0}$; we will use a group closely related to ${\overline {\cal RG}
}_{0}$. So, let us set about defining this auxiliary group. We will
call this auxiliary group ${\overline {\cal RG}}_{0}{}^{\bot}$. First,
consider an element $q$ of ${\overline {\cal RG}}_{0}$. As all the
elements of ${\overline {\cal RG}}_{0}$ are, by definition, homotopic
to the identity, there exists a path $\Gamma$ in ${\overline {\cal
RG}}_{0}$ which connects $q$ with the identity. Consider starting at
the identity end of this path. As the exponential map from $T
{\overline {\cal RG}}_{0}|_{id}$ to a small neighborhood of $id$
is an isomorphism, we may parameterize $\Gamma$ in some small
neighborhood about $id$ with a scalar ${\bar {\epsilon}_1}$
and an element of $T{\overline {\cal RG}}_{0}|_{id}$, $\beta_{1}{}_
{\cal A}$ say. Furthermore, let us assume that $\epsilon_1$
corresponds to the point on $\Gamma$ which is the farthest from
$id$, but yet still within the range of the exponential map's isomorphism
status. Let us denote the element of $\Gamma$ corresponding
to $\epsilon_1$ by $q_1$. Again, we may repeat this construction at
$q_1$ and obtain a point $q_2$ on $\Gamma$ and an element $\beta_{2}
{}_{\cal A}$ of $T{\overline {\cal RG}}_{0}|_{q_1}$ and a scalar $\epsilon_2$.
We can proceed to parameterize all of $\Gamma$ in this manner.
(Note that this will require a finite number of steps, if we choose a
`reasonable' path, as ${\overline {\cal RG}}_{0}$ is compact). So, we
end up with a set $\{ (\epsilon_n,\beta_{n}{}_{\cal A}) \}$ which
parameterizes the path $\Gamma$. However, we can also obtain the
original transformation $q$ by applying the sequence of
infinitesimal transformations given by $\{ (\epsilon_n,\beta_{n}{}_{\cal A})
\}$ to ${\cal M}$, where at the $n^{th}$ step for $p~\epsilon~q_{n-1}({\cal
M})$, $p\rightarrow p +\epsilon_n \beta_n$. Now, let us relate all this to
the auxiliary group ${\overline {\cal RG}}_{0}{}^{\bot}$.

For each element $q$ in ${\overline {\cal RG}}_{0}$, we will define a
corresponding element $q^{\bot}$ in ${\overline {\cal RG}}_{0}{}^{\bot}$ as
follows. Choose a path $\Gamma$ in ${\overline {\cal RG}}_{0}$ as above
and parameterize such a path by the set $\{ (\epsilon_n,\beta_{n}{}_{\cal
A}) \}$. Now, we define $q^{\bot}$ as the transformation on ${\cal M}$
which results from the sequence of transformations $\{ (\epsilon_n,
\beta_{n}{}_{\cal A} \omega^{\cal AB} g_{\cal BC}) \}$. Thus, as one may
trivially see, ${\overline {\cal RG}}_{0}{}^{\bot}$ is compact as a result of
${\overline {\cal RG}}_{0}$'s compactness; also, ${\overline {\cal RG}}_
{0}{}^{\bot}$ is connected by construction. Thus, we will take our group to
be ${\overline {\cal RG}}_{0}{}^{\bot}$, as it satisfies the hypothesis of the
localization theorem and, as we shall see, its Hamiltonian has a nice
interpretation.

So, now let us consider what the function
$\mu(t_\Sigma)$ will be in the case at hand.
As above, we may consider an elment in $T{
\overline {\cal RG}}_{0}{}^{\bot}|_{id}$ and we
have a natural notion of which element to choose,
that corresponding to the beta-function on ${\cal M}$.
Thus, the vector field on ${\cal M}$
which corresponds to this vector in $T{\overline
{\cal RG}}_{0}{}^{\bot}|_{id}$ is simply the vector
field $\beta_{\cal A} \omega^{\cal AB} g_{\cal BC}$. This
vector field is easily shown to have the same fixed
point set as the original beta-function vector field
$\beta_{\cal A}$, as $\omega^{\cal AB}$ and $g_{\cal AB}$
are non-degenerate $\beta_{\cal A}$ and  $\beta_{\cal A}
\omega^{\cal AB} g_{\cal BC}$ both vanish at the same points.
The function on ${\cal M}$ which generates this vector
field via its Hamiltonian flow is $\Psi$, where,

$$
\big( \beta_{\cal C}
\omega^{\cal CB}
g_{\cal BA} \big)
\omega^{\cal AD} =
d\Psi^{\cal D}.
\eqno (5.3)
$$
\vskip 5mm

\noindent Thus, if we note that $\omega_{\cal AB} \omega^{\cal BC}=
\delta_{\cal A} {}^{\cal C}$, then the above equation becomes,

$$
\beta_{\cal A} g^{\cal AB} =
d\Psi^{\cal B}.
\eqno (5.4)
$$
\vskip 5mm

\noindent However, this is essentially the definition of the
Zamolodchikov $C$-Function. For the Zamolodchikov $C$-Function
{\AZ} at any point $p~\epsilon~{\cal M}$ one has $\beta_{\cal B} dc^{\cal B}
= \Omega^{-1} \beta_{\cal A} g^{\cal AB} \beta_{\cal B}$, where $\Omega$
is some normalization constant.  Also, Zamolodchikov proved {\AZ} that
near the fixed point set of $T{\overline {\cal RG}}_{0}{}^{\bot}$ one has
$dc^{\cal B} = \Omega^{-1} \beta_{\cal A} g^{\cal AB}$. Thus, we will
introduce a generalization of the Zamolodchikov $C$-Function, $C$, such
that $dC^{\cal B} = \Omega^{-1} \beta_{\cal A} g^{\cal AB}$ at all $p~
\epsilon~{\cal M}$. As one easily sees, this reduces to the original
Zamolodchikov $C$-Function in the two cases above.
Thus, $\Psi$ is equivalent to the new Zamolodchikov $C$-Function
up to an additive and a multiplicative constant. However, this has some
interesting consequences. If we consider the integral of equation (5.1) in this
case, then we find it takes the following form,

$$
Z =
\int_{\cal M}
DF~
\exp
\Big(
-s ( \Omega dC, \Omega dC )
\Big),
\eqno (5.5)
$$
\vskip 5mm

\noindent where $(~,~)$ is a positive definite invariant quadratic
form on $T^{*}\overline {\cal RG}_{0}{}^{\bot}|_{id}$ and $DF$
is the symplectic measure on ${\cal M}$. However,
this path-integral, as stated previously, is localized
over the set of points at which $dC=0$,  just the set of points
at which $c$ and $C$ agree. But, as is commonly
known \ref \GS{ M. Green, A. Schwarz, and E. Witten
{ \bf Superstring Theory 1: Introduction } (1987) Cambridge
University Press }, see chapter three, requiring the
beta-functions to vanish is equivalent to requiring
that the space-time field's classical stringy equations
of motion are satisfied. Thus, if $\beta_{\cal A} =0$ at some
point $p$, then, for the space-time classical closed string field
theory action $S$, $dS^{\cal A} = 0$ at $p$. From the definition
of the new Zamolodchikov $C$-Function, one sees that if $\beta_
{\cal A}=0$, then $dC=0$, as $g_{\cal AB}$ is non-degenerate. Thus,
the points at which $dC=0$ are exactly the classical solutions of
classical closed string field theory. Furthermore, we may choose the
positive definite invariant quadratic form $(~,~)$ such that
$( \Omega dC, \Omega dC) = \Omega^{2} C^2$. Thus, with this
positive definite invariant quadratic form, we can exactly compute the
path-integral of the field theory defined by $\Omega^2 C^2$. In other
words, we may evaluate,

$$
Z =
\int_{\cal M}
DF~
\exp
\Big(
-s \Omega^2 C^2
\Big),
\eqno (5.6)
$$
\vskip 5mm

\noindent exactly as it reduces to a sum over the
points on ${\cal M}$ which satisfy $dS=0$, the classical
solutions to string field theory. Thus, it seems from the
above argument that the field theory with action $\Omega^2 C^2$ is
a `topological field theory!' I am not really sure of the deeper
meaning of this fact; however, it is rather interesting and
deserves further investigation.

An additional interpretation of the background independent
classical closed string field theory action $S$ is also afforded
by the above exposition. But, we must make some regularity
assumptions on the form of the background independent classical
closed string field theory action, see \ref \MH { M. Henneaux
{\it Lectures on the Anti-Field-BRST Formalism for Gauge
Theories} {\bf Nuclear Physics B } ( Proceedings Supplement )
18A (1990) 47-106 }. We assume that we can split the derivatives
${ {\delta S} / {\delta F^{\cal A}}}$ into $k$ `independent functions'
$I_{\cal A}$ and some `dependent functions' $D_{\cal A}$ in such
a way that the field equations are $I_{\cal A} = 0$, and the form $dI_
{1} \wedge dI_{2} \dots \wedge dI_{k}$ does not vanish on the set of
points at which $dS=0$. Thus, if we make these regularity assumptions
on the action $S$, then we have the following result. Any function $\Phi
(F)$ on ${\cal M}$ which vanishes only on the set $dS=0$ is of the
following form,

$$
\Phi(F) =
Q^{\cal A}
{ {\delta S} \over {\delta F^{\cal A}} },
\eqno (5.7)
$$
\vskip 5mm

\noindent where $Q^{\cal A}$ is a one-form on ${\cal M}$. This result
is rather easy to prove. The regularity conditions that we imposed imply
that the set of `functions' $I_{\cal A}$ can be used as the first $k$
coordinates of a new coordinate system on ${\cal M}$. Thus, if we define
the remaining coordinates to be given by $J_{\cal A}$, then we may write
$\Phi(F)$ in the following form,

$$
\Phi(I,J) =
\Phi(I=0,J) +
\int_{0}^{1}
d \tau
{ { d \Phi(\tau I, J ) } \over { d \tau } }.
\eqno (5.8)
$$
\vskip 5mm

\noindent Thus, we can simplify the above equation by making a change
of variables in the integral and noting $\Phi$ vanishes at $I=0$, to obtain,

$$
\Phi(I,J) =
I_{\cal A}
\int_{0}^{1}
d \tau
{ { d \Phi(\tau I, J ) } \over { d I_{\cal A} } }.
\eqno (5.9)
$$
\vskip 5mm

\noindent Now, if we remember the fact that $I_{\cal A}$ are the field
equations,
one has our result,

$$
\Phi (I,J) =
{ {\delta S} \over {\delta F^{\cal A} } }
\int_{0}^{1}
d \tau
{ {\delta \Phi(\tau I,J) } \over {\delta I_{\cal A}} },
\eqno (5.10)
$$
\vskip 5mm

\noindent were we have called the integral above $Q^{\cal A}$. Thus,
applying this to the case at hand, we find that, as the `functions' $dC^
{\cal A}$ vanish only on ${\cal M}$ at the points $dS=0$, $dC^{\cal A}$
must be of the following form,

$$
d C^{\cal A} =
Q^{\cal A}_{\cal B}
dS^{\cal B},
\eqno (5.11)
$$
\vskip 5mm

\noindent where $Q^{\cal A}_{\cal B}$ is a vector valued one-form over
${\cal M}$ given by,

$$
Q^{\cal A}_{\cal B} \equiv
\int_{0}^{1}
d \tau
{ { {\delta}^2 C(\tau I,J) } \over
{ \delta I_{\cal A}  \delta F^{\cal B} } },
\eqno (5.12)
$$
\vskip 5mm

\noindent where $F$ runs over both $I$ and $J$. Thus, if we assume all
of the equations of motion of string field theory are independent, we have
a natural relation between the new Zamolodchikov $C$-Function and the
background independent classical closed string field theory action $S$,

$$
dC^{\cal A} =
dS^{\cal B}
\int_{0}^{1}
d \tau
{ { {\delta}^2 C(\tau F) } \over
{ \delta F_{\cal A}  \delta F^{\cal B} } }
\eqno (5.13)
$$
\vskip 5mm

\noindent Again, what this means? However, it does lead to a
new and novel interpretation of the action $S$. Indeed, if $Q^{\cal A}_
{\cal B}$ is invertible, then one has, up to a physically irrelevant
constant,

$$
S =
\int
{Q^{-1}}_{\cal A}^{\cal B}
dC^{\cal A}~
DF_{\cal B}.
\eqno (5.14)
$$
\vskip 5mm

\noindent Which expresses the background independent classical
closed string field theory action $S$ completely in terms of the new
Zamolodchikov $C$-Function!

\newsec {Conclusions}

In this article we've proved that the field theory action $C^2$ yields a
`topological field theory' over ${\cal M}$, the space of all
two-dimensional field theories. Also, we've proved that the
background independent classical closed string field theory action $S$
is intimately related to the new Zamolodchikov $C$-function. (However,
with both we've assumed ${\omega_{\cal AB}}$'s existence). Thus, with
these new results we should be able to gain some insights into
background independent closed string field theory. As the theory
with the action $C$ is the `square-root' of a `topological field' theory
and $C$  is so intimately related to $S$, we should probably think
of $S$ as the `square-root' of the `topological field theory' $C^2$.
What this means ?

\bigbreak\bigskip\bigskip\centerline{{\bf Acknowledgments}}\nobreak
I would like to thank Elizabeth Fraser, Heidi Berry, Warren Defever, Mark
Kozelek, Lisa Gerrard, Brendan Perry, and 4AD for the music by which this was
written, and for the music which is this article's motivation. Also, I would
like to thank Jessica Mendels for her support.

\listrefs
\bye